\documentclass[showpacs,aps,graphicx,twocolumn]{revtex4}
\usepackage{verbatim}
\usepackage{graphicx}
\usepackage{algorithm}
\usepackage{algorithmic}
\usepackage{subfigure}
\usepackage{multirow}
\usepackage[caption=false,font=scriptsize,labelfont=sf,textfont=sf]{subfig}
\usepackage{color}

\begin{document}
\title{Evolutionary-based quantum architecture search}

\author{Anqi Zhang, Shengmei Zhao\footnote{
		Email address:zhaosm@njupt.edu.cn}}
\address{
Institute of Signal Processing Transmission, Nanjing University of Posts and Telecommunications (NUPT), Nanjing, 210003,  China \\
}
\date{\today }

\begin{abstract}
Quantum architecture search (QAS) is desired to construct a powerful and general QAS platform which
can significantly accelerate quantum advantages in error-prone and depth limited quantum circuits in today Noisy Intermediate-Scale Quantum (NISQ) era.
In this paper, we propose an evolutionary-based quantum architecture search (EQAS) scheme for the optimal layout to balance the higher expressive power and the trainable ability.
In EQAS, each layout of quantum circuits, i.e quantum circuit architecture(QCA), is first encoded into a binary string, which is called quantum genes later. Then, an algorithm to remove the redundant parameters in QCA is performed according to the eigenvalues of the corresponding quantum Fisher information matrix (QFIM). Later, each QCA is evaluated by the normalized fitness, so that the sampling rate could be obtained to sample the parent generation by the Roulette Wheel selection strategy. Thereafter, the mutation and crossover are applied to get the next generation.
EQAS is verified by the classification task in quantum machine learning for three datasets. The results show that the proposed EQAS can search for the optimal QCA with less parameterized gates, and the higher accuracies are obtained by adopting EQAS for the classification tasks over three dataset.
\end{abstract}
\maketitle

\section{Introduction}
The parameterized quantum circuits are the essential components in the variational quantum algorithms (VQAs) operated in the Noisy Intermediate-Scale Quantum(NISQ) devices\cite{mcclean2018barren,bruzewicz2019trapped,krantz2019quantum}. The effectiveness of VQAs with small-scale structure has been demonstrated by many works\cite{cai2019once,guo2020single,peruzzo2014variational,versluis2017scalable,tannu2019not,murali2019noise,murali2020software,wu2021tilt,zhang2022hidden,li2019tackling,ding2020systematic}. It is shown that the performance of VQAs degrades significantly with bigger number qubit and larger circuit depth, which is caused by the trade-off between the expressivity and the trainability. Therefore, optimizing the structure of the quantum circuits with low depth and less qubits is one main challenge in realizing VQAs.

Compared with the quantum circuit with a fixed structure for VQAs\cite{schuld2020circuit,chen2021end,bhatia2019matrix,adhikary2021entanglement,zhang2022quantum,perez2020data}, a variable structured quantum circuit is a promising way, since this structure can not only maintain a smaller depth to suppress noise caused by the imperfect quantum gates, but also can keep sufficient expressive power to implement the optimizing tasks. For example,
Oddi et al.\cite{hart1987semi} proposed a search method to obtain the nearest-neighbor quantum circuits based on Greedy Randomized algorithm.
Zhang et al.\cite{liu2018darts,zhang2020differentiable} introduced an automated search approach to get the quantum circuits architecture based on differentiable architecture searching.
Chen et al.\cite{kuo2021quantum,ye2021quantum} presented a quantum circuit architecture scheme based on Deep Reinforcement Learning.
Liu et al.\cite{yao2022monte,meng2021quantum} proposed a strategy to find the optimal layout of the quantum circuit based on Monte Carlo Tree Search.
And D.Szwarcman et al.\cite{szwarcman2019quantum,szwarcman2022quantum,ye2020quantum,wang2022quantumnas} presented a quantum-inspired search algorithm for the deep networks based on the evolutionary or genetic algorithms.

Since different parameterized quantum gates play different roles in quantum circuits \cite{meyer2021fisher,abbas2021power,jha2022bayesian}, it is possible to remove some less important parameterized quantum gates without affecting its expressive power, so that the efficiency of the quantum circuits can be improved. For example, a quantum circuit pruning was presented in \cite{wang2022quantumnas} to remove the quantum gates whose parameter values are close to zero, however,
the loss function of the optimization task may be fail to converge by its brute-force strategy. 
For a small change in some important parameters can cause a significant change of its quantum state, therefore,
how to efficiently find the quantum circuit structure with best performance is still an open question.

In this paper, we propose an evolutionary-based quantum architecture search scheme, named EQAS, to obtain the optimal layout of quantum circuits, and it is called quantum circuit architecture(QCA) for quantum optimization tasks. In the scheme, the QCAs are encoded into binary strings as quantum genes, which are randomly generated from the operation pool with quantum gates at beginning. The quantum Fisher information matrix(QFIM) of each  quantum gene is then calculated, and the quantum gates containing redundant parameters in the QCA are removed according to the eigenvalues of QFIM. The performance of QCA, together with the length of the QCA, are evaluated to derive its fitness. The normalized fitness values are then used as the probabilities to sample the parent generation by using the Roulette-wheel selection strategy, and the mutation and crossover are later operated on the quantum genes to get the children generation. The above steps are repeated until a QCA with satisfactory performance is found or the number of iterations is reached. Finally, we testify the proposed scheme with the classification tasks on three different datasets.

The advantages of the proposed searching scheme are three-fold: 1)We propose an evolutionary-based quantum architecture search, named EQAS, to obtain the optimal layout of quantum circuits. 2)We use QFIM to mark and remove the redundant parameters in one QCA. 3)The proposed EQAS scheme is demonstrated to have a good performance with a less parameter for the classification tasks.

The paper is organized as follows. In Sec.II, we present the details of  EQAS for finding the optimal QCA. In Sec.III, we discuss the simulation results for the classification tasks in Iris, MNIST and MNIST-Fashion datasets. Finally, we draw the conclusion in Sec.IV.

\section{The evolutionary-based searching method for QCA}
\begin{figure*}[!htpb]
	\centering
	\includegraphics[width=1\textwidth]{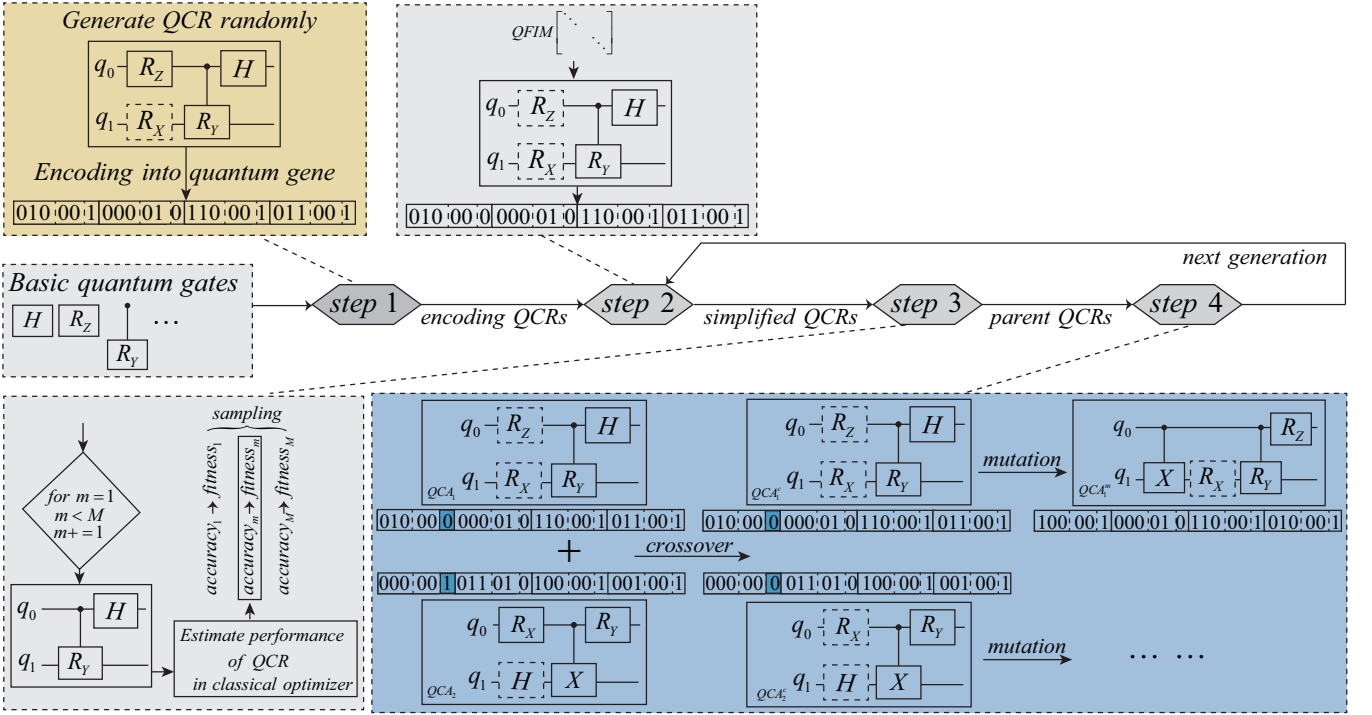}
	\caption{The framework of the EQAS scheme for QCA. In step(1), a set of QCAs is generated randomly and then encoded into quantum genes which are binary strings.
	In step(2), the QFIM of each QCA is calculated and diagonalised to simplify the QCA by removing redundant parameters based on the variations in the eigenvalues of the QFIM.
	In step(3), the performance of QCAs is evaluated in a Adam-based optimizer hybrid quantum-classical method and the value of fitness is calculated according to the accuracy and the number of quantum gates used in the QCA. Roulette-wheel selection is carried out to get the parent generation of QCAs.
	In step(4), crossover and mutation are applied on the parent generation to get the offspring.
	Steps (2)-(4) are iterative until one QCA meets the stop condition or the number of iterations reaches the upper limit.}
	\label{Fig:1}
\end{figure*}

The framework of the EQAS scheme for QCA is described in FIG.~\ref{Fig:1}. Firstly, a set of QCAs generated randomly are encoded as binary strings termed quantum genes, where each quantum gene has three segments, corresponding to three gene information, such as, the "TYPE", the "PLACE" and the  "INCLUDED". Next, the QFIM of each QCA is calculated and diagonalised. Based on the QFIM,  the redundant parameter of a QCA is located and then removed, until the eigenvalues of that QCA's QFIM are all non-zero.
Then, the accuracy performance of each quantum gene is estimated by
the Adam-based optimizer hybrid quantum-classical method. The fitness value of each QCA is calculated according to the accuracy and the number of used quantum gates in the QCA.
Roulette-wheel election strategy is adopted to sample the parent population according to the normalized fitness values. Finally, the crossover and mutation operations are carried out to generate the children generation. 

\subsection{Quantum genes of QCAs}
A quantum circuit can be composed by some single-qubit gates or double-qubit gates. Here, 8 types quantum gates are used, in which $R_x,R_y,R_z,H$ are the single-qubit gates and $C_{not},CR_x,CR_y,CR_z$ are the double-qubit gates. And these qubit gates are defined an operation pool.

\begin{table}[!htpb]
	\caption{Encoding of quantum genes for QCA.}
	\begin{ruledtabular}
		\begin{tabular}{ccccccccc}
			\hline
			\multirow{2}{*}{TYPE} &000&001&010&011&100&101&110&111  \\
			\cline{2-9}
			&$R_X$&$R_Y$&$R_Z$&$H$&$C_{not}$&$CR_X$&$CR_Y$&$CR_Z$\\
			\hline
			\multirow{2}{*}{PLACE} & \multicolumn{2}{c}{00} &\multicolumn{2}{c}{01}
			& \multicolumn{2}{c}{10} &\multicolumn{2}{c}{11} \\
			\cline{2-9}
			& \multicolumn{2}{c}{$q_0$} &\multicolumn{2}{c}{$q_1$}
			& \multicolumn{2}{c}{$q_2$} &\multicolumn{2}{c}{$q_3$} \\
			\hline
			\multirow{2}{*}{INCLUDED} & \multicolumn{4}{c}{1} &\multicolumn{4}{c}{0}  \\
			\cline{2-9}
			& \multicolumn{4}{c}{Dominant} &\multicolumn{4}{c}{Recessive}  \\
		\end{tabular}
		\label{tab:1}  
	\end{ruledtabular}
\end{table}

The architecture of a QCA can be represented by the selection of quantum gates from the operation pool and the layout of these gates in a quantum circuit.
Here, the QCA is constructed by building blocks of quantum gates that have the same layouts and each block contains one single-qubit gate layer and one double-qubit gate layer.

After fixing the number of qubits used in the quantum circuit, we randomly select the same number of single-qubit gates and double-qubit gates from the operation pool at first. Next, we apply the single-qubit gates on each qubit in sequence, and place the double-qubit gates  sequentially on each qubit in a circle order, that is, the controlled qubit is always the next qubit of the control qubit.
As shown in Table.~\ref{tab:1}, each quantum gate is represented by three bits with the segment "TYPE". "PLACE" is the segment to describe where the quantum gate applies on, and "INCLUDED" segment represents  whether the quantum gate works in the quantum circuit. Here, 1 is the dominant(working) and 0 is the recessive(no-working). Each quantum gate is encoded to a quantum gene with binary bits, and a QCA is then encoded to a unique binary string. 	
For example, as shown in the 'step 1' of FIG.~\ref{Fig:1}, a QCA with four quantum gates which operated on a 2-qubit quantum circuit is randomly generated.
Hence, four quantum genes are encoded. The first three bits of the quantum genes are $010,000,110,011$, denoting  $R_z,R_x,CR_y,Hadamard$; The following two bits of the quantum genes are $00,01,00,00$, representing $R_z,CR_y,Hadamard$  applied on the qubit $q_0$ and $R_x$ applied on the qubit $q_1$; The final one bit of the quantum genes are $1,0,1,1$, denoting that $R_z,CR_y,Hadamard$ are truly working in this quantum circuit, while $R_x$ does not. For the double-qubit gate, the "PLACE" segment only marks the control qubit. Therefore, the control qubit of $CR_y$ is $q_0$ and the controlled qubit is $q_1$.

\subsection{QFIM-based algorithm}
The capacity of a parametrized quantum circuit(PQC) can be expressed by the effective dimension of the PQC \cite{thrun2004advances,haug2021capacity}, which is equals to the total number of the independent parameters in a quantum circuit with fixed structure. Therefore, the QCA can be simplified when the redundant parameters are removed while the expressive power is kept or the variation of the quantum state caused by the removed redundant parameters is small. In other words, the redundant parameters of a QCA  do little contribution on the changing of the quantum state.

According to Ref.\cite{thrun2004advances,haug2021capacity}, the parameters corresponding to the change of the number of 0 eigenvalue of QFIM can be removed from the QCA. Hence, we delete a parameterized quantum gate according to the following steps.  (1) we mark all parameterized quantum gates and estimate the number of 0 eigenvalues of their QFIM; (2) we sequentially take a parameterized quantum gate out from the QCA one by one, and estimate the number of 0 eigenvalues of QFIM for the left QCA. We delete the parameterized quantum gate when there is a change in the number of 0 eigenvalues in step1 and step 2. The process is continued until there is non-zero eigenvalue for the left QCA. We show the details at Algorithm.~\ref{alg1}.

\begin{algorithm}
	\caption{Removing algorithm for redundant parameters of the QCA algorithm}
	\label{alg1}
	\begin{algorithmic}[1]
		\REQUIRE QCA $C$; number of parameters $N$; parameters with assigned serial number $p_k,k=1\dots N$; QFIM $F(\theta_{random})$; the rank of QFIM $r$; an empty set $K=\{\}$;
		\STATE Diagonalise $F(\theta_{random})$ and count the number of eigenvalues with a value of 0, $E$. $k=1.$
		\WHILE{$F(\theta_{random)}$ has zero eigenvalues}
		\STATE{$\qquad$Remove the parameter $p_k$, and update $F(\theta_{random})$ by deleting the row $k$ and the column $k$;}
		\STATE{$\qquad$Re-diagonalize $F(\theta_{random})$ and count the number of eigenvalues with a value of 0, $E_{k}^{'}$.}
		\IF{$E_{k}^{'} < E$}
		\STATE {$\qquad$Add $k$ to the set $K$;}
		\STATE {$\qquad$$E = E_{k}^{'}$.}
		\ELSE
		\STATE {$\qquad$Remain $E$;}
		\STATE {$\qquad$$k=k+1$.}
		\ENDIF 	
		\ENDWHILE	
		\STATE { Removing parameters in the set $K$ from $C$}
	\end{algorithmic}
\end{algorithm}

\subsection{Fitness and Sampling}
In this subsection, we discuss how to calculate the fitness and we also present the sampling strategy.

\textbf{Fitness}
The fitness can be calculated by $Acc_{QCA}$ and the length of QCA $Len_{QCA}$, i.e. the number of quantum gates used in the QCA. The expression is
\begin{equation}\label{eqn3}
	fitness = (1-\alpha)Acc_{QCA} + \alpha \times\frac{1}{Len_{QCA}} ,
\end{equation}
where $\alpha$ is a weight that balances the accuracy and the length,
$Acc_{QCA}$ is the final accuracy of the QCA by minimizing the cost function given in the specific optimization task.

\begin{figure}[t]
	\includegraphics[scale=1.2]{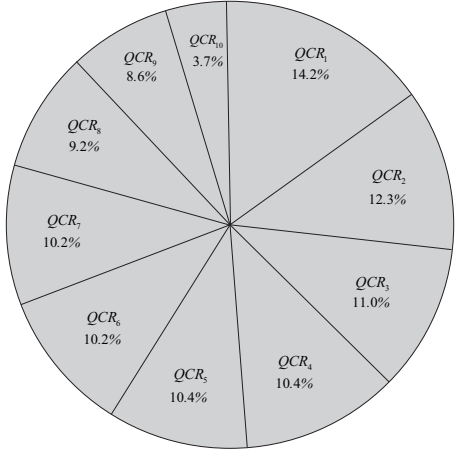}
	\caption{Sampling strategy of Roulette-wheel selection. The probability of each $QCA_i$ being sampled was calculated by fitness, and $QCA_1$ with the high probability is more likely to be extracted while $QCA_{10}$ is not.
		\label{fig:2}}
\end{figure}

\textbf{Sampling strategy} Roulette-wheel selection strategy\cite{lipowski2012roulette} is adopted to sample the parent population, as shown in FIG.~\ref{fig:2}.
We normalize the fitness obtained in the above step, and sample the QCAs corresponding to the normalized fitness in which
the normalized value is used as the probability of its corresponding QCA, until the number of samples reaches the size of the population. For example, there are 10 QCAs in FIG.~\ref{fig:2}, the fitness are 0.23, 0.54, 0.58, 0.64, 0.64, 0.65, 0.65, 0.69, 0.77, and 0.89 respectively. Then the normalized fitness, that is also the probability of each QCA, are $3.7\%$, $8.6\%$, $9.2\%$, $10.2\%$, $10.2\%$, $10.4\%$, $10.4\%$, $11.0\%$, $12.3\%$, and $14.2\%$, respectively. With the highest probability,  $QCA_1$ is then more likely sampled out than $QCA_{10}$.

\subsection{Crossover and mutation}
In this subsection, we present the crossover and mutation strategy.

\begin{figure}[t]
	\includegraphics[width=.44\textwidth]{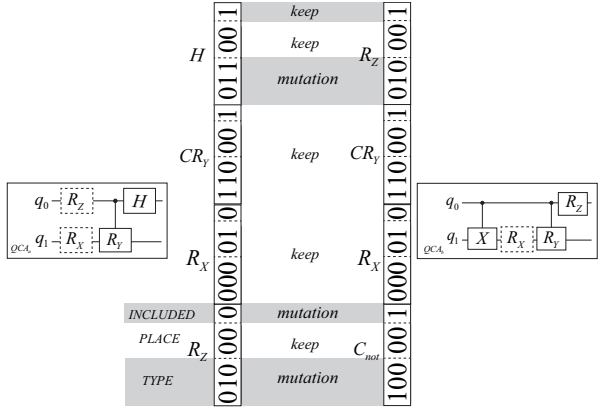}
	\caption{The details of mutation.
		\label{fig:3}}
\end{figure}

\textbf{Crossover}
If both the "INCLUDED" segments of the two-parent QCAs at the same position are 1, then the corresponding segments at this position for their offspring are retained; Otherwise the corresponding segments at this position for their offspring are both set to 0.
Therefore, two-parent QCAs produce two offspring who have the same number of quantum gates with dominant "INCLUDED".
As shown in the 'step 4' of FIG.~\ref{Fig:1}, $QCA_1$ and $QCA_2$ have different "INCLUDED" segments in the first quantum gene(the left dark blue box). Therefore, the "INCLUDED" segment of $QCA_1$ and $QCA_2$ are both set to 0(the right dark blue box) during crossover. The other three genes who have the same "INCLUDED" segments at the same position are retained in their offsprings.

\textbf{Mutation} In mutation, the bits in "PLACE" segment of a gene is kept, while the bits in "TYPE" and "INCLUDED" segments are mutated according to the set probability. Hence, the mutation changes the type and the number of the quantum gates. 
As shown in FIG.~\ref{fig:3}, the four "PLACE" gene bits of the $QCA_{a}$ are unchanged, while the gene bits in "TYPE" and "INCLUDED" segments are  changed with the set probability. After mutation, the $R_z$ gate with dominant "INCLUDED" in the $q_0$ position of $QCA_{a}$ becomes the $C_{not}$ gate in the $QCA_{b}$, the $Hadamard$ gate in the $q_0$ position becomes the $R_{z}$ gate in the $QCA_{b}$. The $R_x$ gate in the $q_1$ position and the $CR_y$ gate in the $q_0$ position of $QCA_{a}$ are unchanged in the $QCA_{b}$.

\section{Results}

\begin{table*}[!htpb]
	\caption{Results of simulation by using the EQAS and the Q-NAS for three classification tasks.}
	\begin{ruledtabular}
		\begin{tabular}{cccccc}
			\hline
			\multirow{1}{*}{}  &Dataset&Search method&Accuracy&Gates&Para gates  \\
			\hline
			\multirow{2}{*}{2-qubit}&   \multirow{2}{*}{Iris-(Setosa VS. Versicolour)}
			&EQAS&100\%&1&1  \\
			\cline{3-6}
			& &Q-NAS&100\%&2&2  \\				
			\hline		
			\multirow{4}{*}{4-qubit} &\multirow{2}{*}{MNIST-(3 VS. 6)}&EQAS&94\%&22&9  \\
			\cline{3-6}
			& &Q-NAS&96\%&24&22  \\
			
			\cline{2-6}              &\multirow{2}{*}{Fashion-MNIST-(Dress VS. Shirt)}&EQAS&75\%&24&20 \\
			\cline{3-6}
			& &Q-NAS&78\%&24&23  \\
			\hline
		\end{tabular}
		\label{tab:2}  
	\end{ruledtabular}
\end{table*}

\begin{figure*}[t]
	\subfigure[] {\includegraphics[width=.2\textwidth]{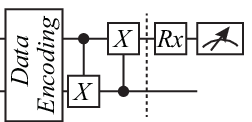}}
	\subfigure[] {\includegraphics[width=.65\textwidth]{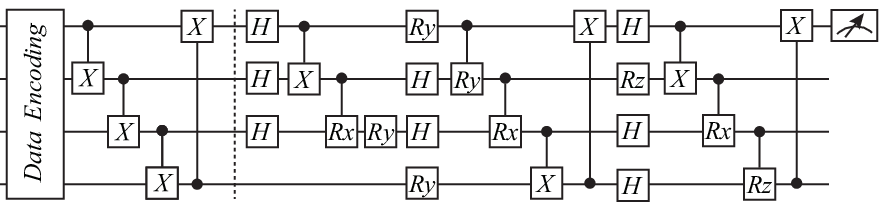}}
	\subfigure[] {\includegraphics[width=.65\textwidth]{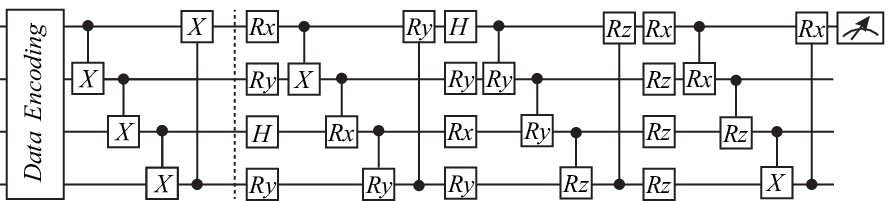}}
	\caption{The quantum circuits of the final searched QCA after 100 iterations. (a) for Iris, (b) for MNIST and (c) for Fahsion-MNIST. After encoding training data into the quantum circuit, a circle of $C_{not}$s is applied to connect all qubits.
		\label{fig:8}}
\end{figure*}

\begin{figure}[t]
	\includegraphics[width=.47\textwidth]{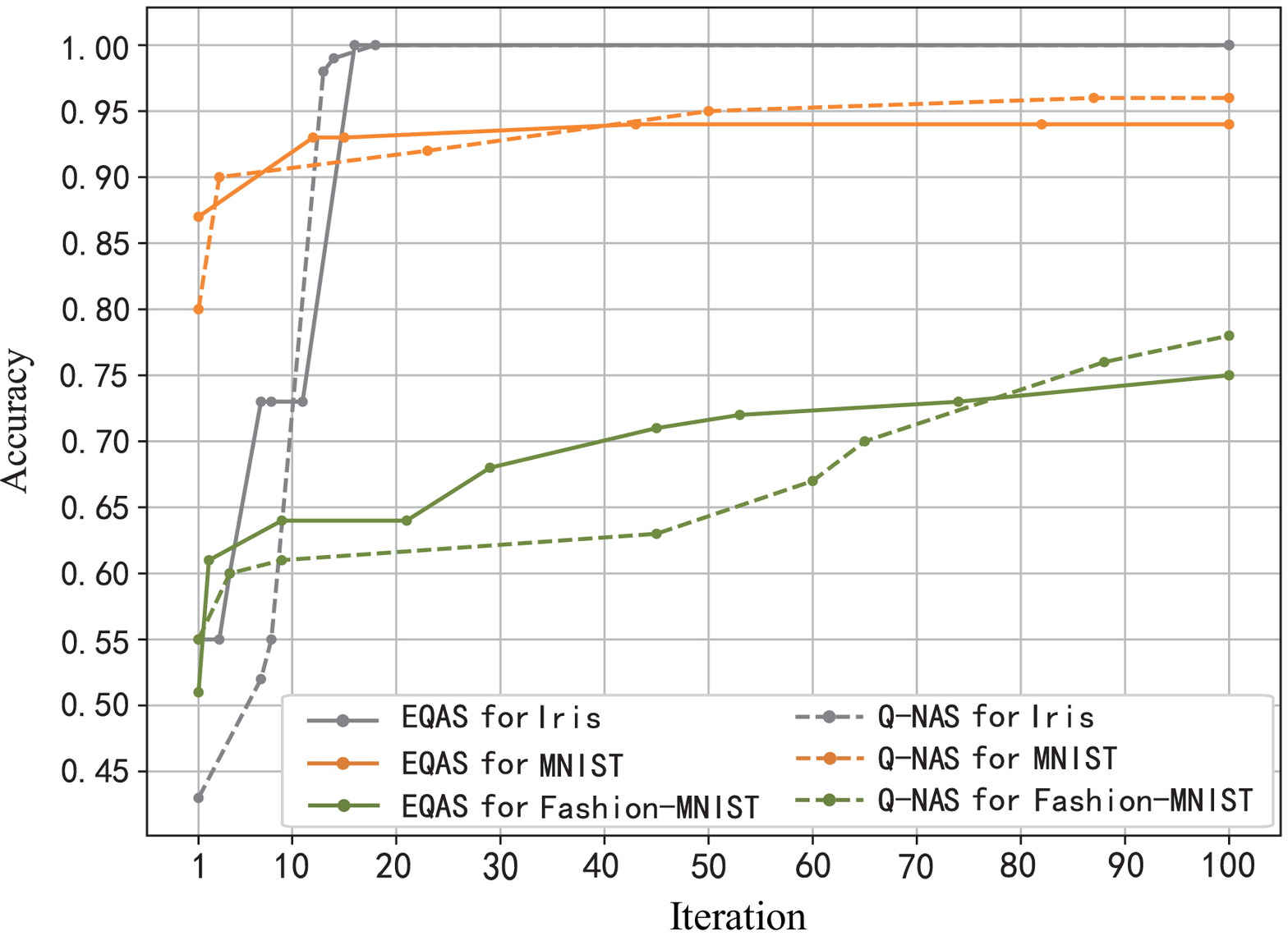}
	\caption{The classification accuracy against the iterations for the Iris(grey), MNIST(yellow) and Fashion-MNIST(green) datasets.
		\label{fig:9}}
\end{figure}
We demonstrate the numerical simulations of the proposed scheme on 3 classification datasets. They are, Iris (Setosa, Versicolour); MNIST-10\cite{lecun1998gradient} (3, 6);
and Fashion-MNIST\cite{xiao2017fashion} (dress, shirt).
The Iris images are 4-dimension data and encoded by the method of angle encoding\cite{schuld2021quantum} into 2-qubit QCA with 5 quantum gates initialized in the QCA; the MNIST and Fashion-MNIST images are center-cropped and down-sampled from $28 \times 28$ to $8\times8$ for encoding by amplitude encoding\cite{schuld2021quantum} into 4-qubit QCA with 3 blocks (one block with a 4-single-qubit-gate layer and a 4-double-qubit-gate layer) initialized in the QCA. 
Here, $C_{not}$ gate is adopted to connect the neighbouring qubit.
For all the searched QCAs, the same training setting are applied. Additionally, Adam optimizer with initial learning rate 0.1 is adopted in this work. The batch size is 30 for the task with 4-qubit QCA searching, and the batch size is 10 for 2-qubit QCAs searching. The iterations is 100, the population size is 30 and the mutation probability is 0.4.

The results are shown in Table.~\ref{tab:2}, together with the comparison with Q-NAS proposed in \cite{wang2022quantumnas}. In Iris classification task, the EQAS scheme reduces the number of quantum gates by 80\% during searching processing from the initial 5 quantum gates, to finally 1 quantum parameterized gate for 100\% accuracy, the searched QCA is shown as FIG.~\ref{fig:8}(a), there is only one quantum parameterized gate left for the task. While the Q-NAS achieves 100\% accuracy by 2 quantum parameterized gates; 
For MNIST classification task, the EQAS scheme reduces the number of quantum gates by 8.3\% during searching processing, and finally achieves 94\% accuracy, the quantum circuit of the searched QCA is shown as FIG.~\ref{fig:8}(b), where the final quantum circuits has 22 quantum gates included 9 parameterized gates. For the same task, the Q-NAS achieves 96\% accuracy by 24 quantum parameterized gates included 22 parameterized gates; The quantum parameterized gates used in MNIST classification task are apparently reduced by using the proposed EQAS scheme.
For Fashion-MNIST classification task, the EQAS scheme achieves 75\% accuracy by 24 quantum parameterized gates included 20 parameterized gates, and the quantum circuit of the searched QCA is shown as FIG.~\ref{fig:8}(c). While the Q-NAS achieves 78\% accuracy by 24 quantum parameterized gates included 23 parameterized gates. The number of the used quantum parameterized gates is reduced by using the proposed EQAS scheme. 
Compared by the Q-NAS, the proposed EQAS scheme can use less parameterized gates, even less quantum gates, to complete the classification tasks, that is, the EQAS scheme can reduce the number of quantum gates used in quantum circuit.

FIG.~\ref{fig:9} shows the classification accuracy against the iterations for the three classification tasks by using the proposed EQAS scheme and the Q-NAS scheme. The results show that the classification accuracies are approaching to the final accuracies with the increase of the iteration both for the proposed EQAS scheme and the Q-NAS scheme. The proposed EQAS scheme can reach a high accuracy faster than the Q-NAS scheme. But the Q-NAS can achieve a higher accuracy at the end of iterations because the Q-NAS scheme has more quantum gates used in the quantum circuits.

\section{Conclusion}
We have proposed an EQAS scheme for searching the optimal layout of the quantum circuits to balance the higher expressive power and the trainable ability. Each quantum circuit architecture(QCA) has encoded into a bit string consisting of some quantum genes, and the redundant parameterized gates inside have been removed by a designed algorithm based on QFIM. QCA has then evaluated by an Adam-based hybrid quantum-classical, and their normalized fitness have been used as the sampling probabilities to sample the parent generation by the Roulette Wheel selection strategy. Finally, the mutation and crossover have been applied to get the next generation. 
The EQAS scheme has been verified to search for the optimal layout of QCAs for classifying tasks in quantum machine learning with Iris, MNIST and Fashion-MNIST datasets. 
Compared with Q-NAS scheme, the proposed EQAS scheme can obtain a higher classification accuracy for the three datasets with the QCAs having a less quantum parameterized gates, even a smaller number quantum gates.
With the advantage of efficient selecting the optimal QCA and the hybrid quantum-classical, the proposed EQAS can be used to solve others tasks, such as the variational quantum eigensolver (VQE), at the Noisy Intermediate Scale Quantum (NISQ) device era.

\section*{Appendix A: Performances of Algorithm.~\ref{alg1}.}
\setcounter{equation}{0}
\setcounter{subsection}{0}
\renewcommand{\theequation}{B.\arabic{equation}}

\begin{figure*}[t]
	\subfigure[] {\includegraphics[width=.49\textwidth]{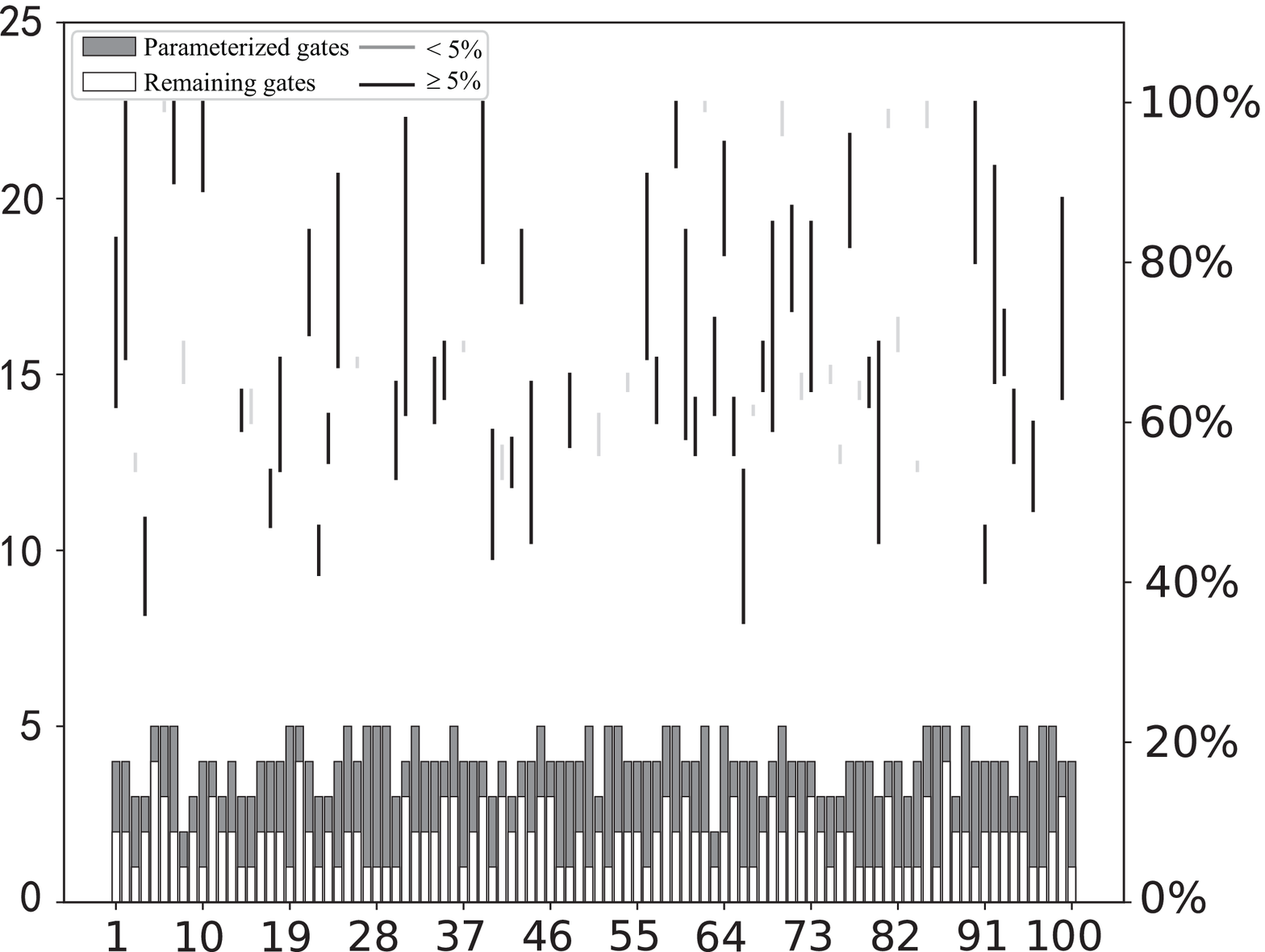}}
	\subfigure[] {\includegraphics[width=.49\textwidth]{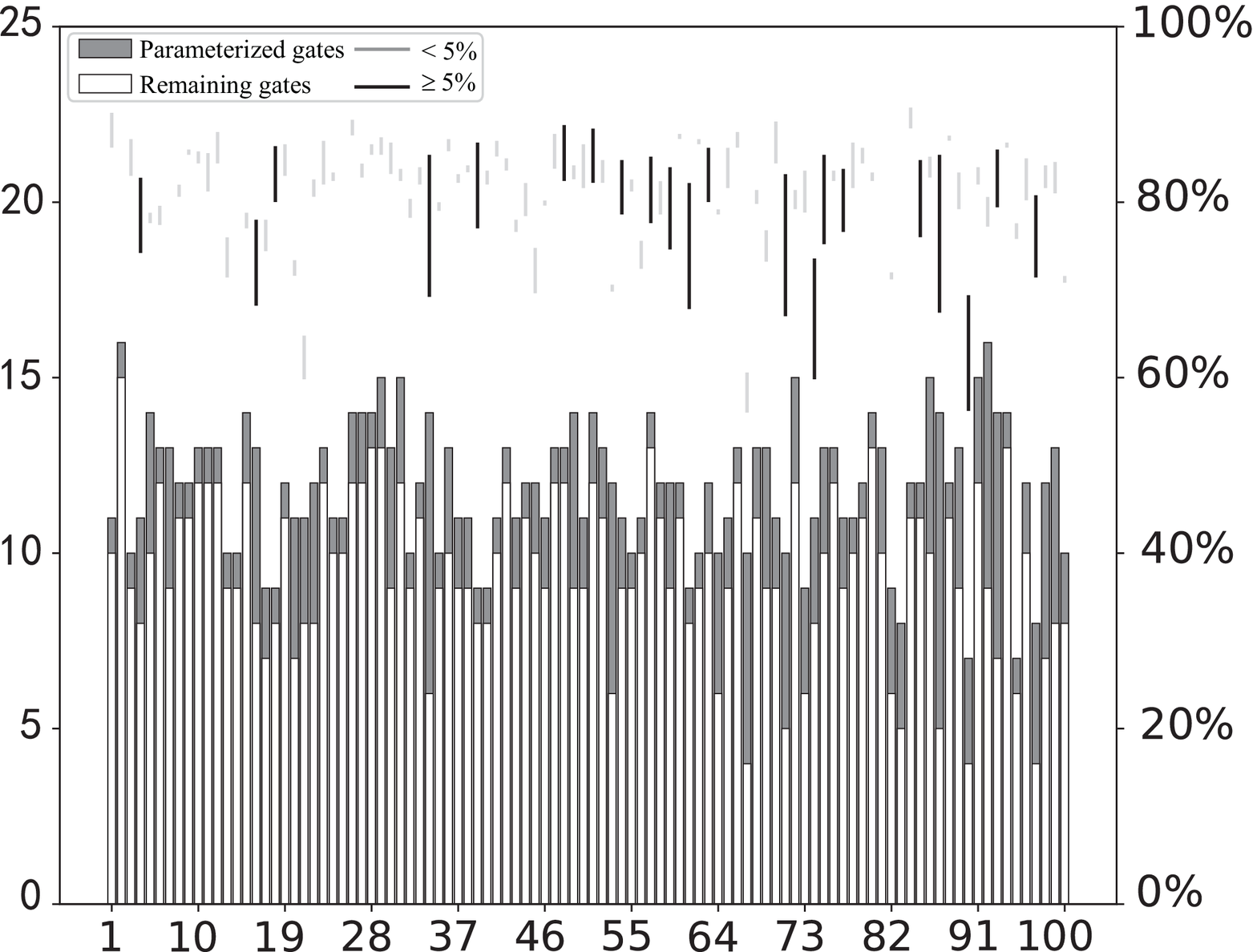}}
	\caption{The performance of the Algorithm.~\ref{alg1} applying on QCAs with 2-qubit(for the Iris dataset) and 4-qubit(for the MNIST dataset) framework.
		\label{fig:6}}
\end{figure*}

In the EQAS scheme, quantum gates with redundant parameters are removed by using the knowledge of QFI and the steps of removing parameterized quantum gates with redundant parameters from QCA are described in the Algorithm.~\ref{alg1}. We show the performances of QCAs that remove the quantum gates with redundant parameters or not in FIG.~\ref{fig:6}. 100 2-qubit QCAs for the Iris dataset in FIG.~\ref{fig:6}(a) and 100 4-qubit QCAs for the MNIST in FIG.~\ref{fig:6}(b) are randomly generated. The height of each bar represents the number of parameterized quantum gates used in each QCA, where the black bar represents the number of parameterized quantum gates used in each QCA, and the white bar represents the number of parameterized quantum gates left after removing the redundant parameters. Each bar has a vertical line above it, representing the change in accuracy after removing the redundant parameterized quantum gates. The longer the vertical line, the greater the change in accuracy. Here, we only care about the range of accuracy variation rather than the direction(lower or higher). The dark vertical lines represent that the variation of accuracy is more than 5\%; while the light lines represent less than 5\%.

FIG.~\ref{fig:6}(a) shows the case of the Iris dataset using QCAs of the 2-qubit framework with 5 quantum gates at most. Among them, 45 QCAs changed their accuracy by more than 5\% after removing redundant parameterized gates. FIG.~\ref{fig:6}(b) shows the case of the MNIST dataset using QCAs of the 4-qubit framework with 2 blocks of quantum gates(16 quantum gates) at most and a total of 19 QCAs changed their accuracy by more than 5\%. Compared FIG.~\ref{fig:6}(a) with  FIG.~\ref{fig:6}(b), the redundant parameter removing method of Algorithm.~\ref{alg1} is more suitable for QCAs with more parameterized quantum gates. The reason is that the redundant parameters removed by QFI do not completely have no influence on QCA, but have lesser influence than other parameters, so the Algorithm.~\ref{alg1} is more suitable for QCA with a large number of quantum gates.

\section*{Appendix B: Performances of the Roulette-wheel selection and the Elite selection.}
\setcounter{equation}{2}
\setcounter{subsection}{2}
\renewcommand{\theequation}{B.\arabic{equation}}
The EQAS scheme generates the offspring through three basic operations, selection, crossover, and mutation. The next generations with low fitness are gradually eliminated and those with high fitness are kept in every iteration. In this way, after $N$ generations, it is possible to evolve individuals with high fitness.

\begin{figure}[t]
	\includegraphics[scale=0.6]{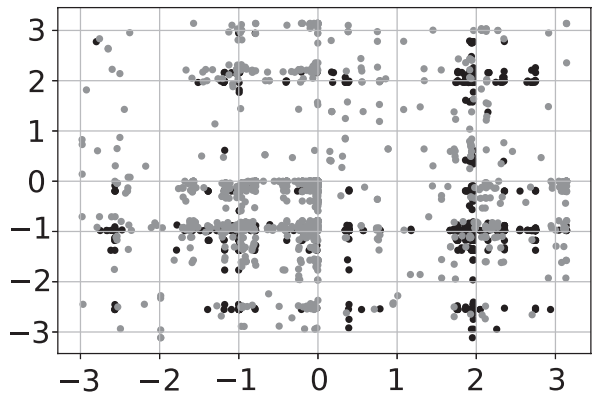}
	\caption{Performances of the Roulette-wheel selection(grey circles) and the Elite strategy(black circles).
		\label{fig:7}}
\end{figure}

The parent population needs to meet two characteristics, high fitness that can produce offspring with better performance in a greater probability, and enough diversity that prevents from generating very similar offspring. FIG.~\ref{fig:7} Shows the results of using two strategies of selection, Roulette-wheel selection(grey circles) and Elitist election(black circles), to produce a parent population, and then mutating and crossing to produce the corresponding offspring.

In FIG.~\ref{fig:7}, the grey circles cover a larger area than the black circles, which indicates the method of Roulette-wheel selection produces more different offspring than Elitist election.

\section*{Appendix C: The final searched QCA for the three datasets.}
\setcounter{equation}{0}
\setcounter{subsection}{0}
\renewcommand{\theequation}{B.\arabic{equation}}

\begin{figure*}[!htpb]
	\subfigure[] {\includegraphics[width=.31\textwidth]{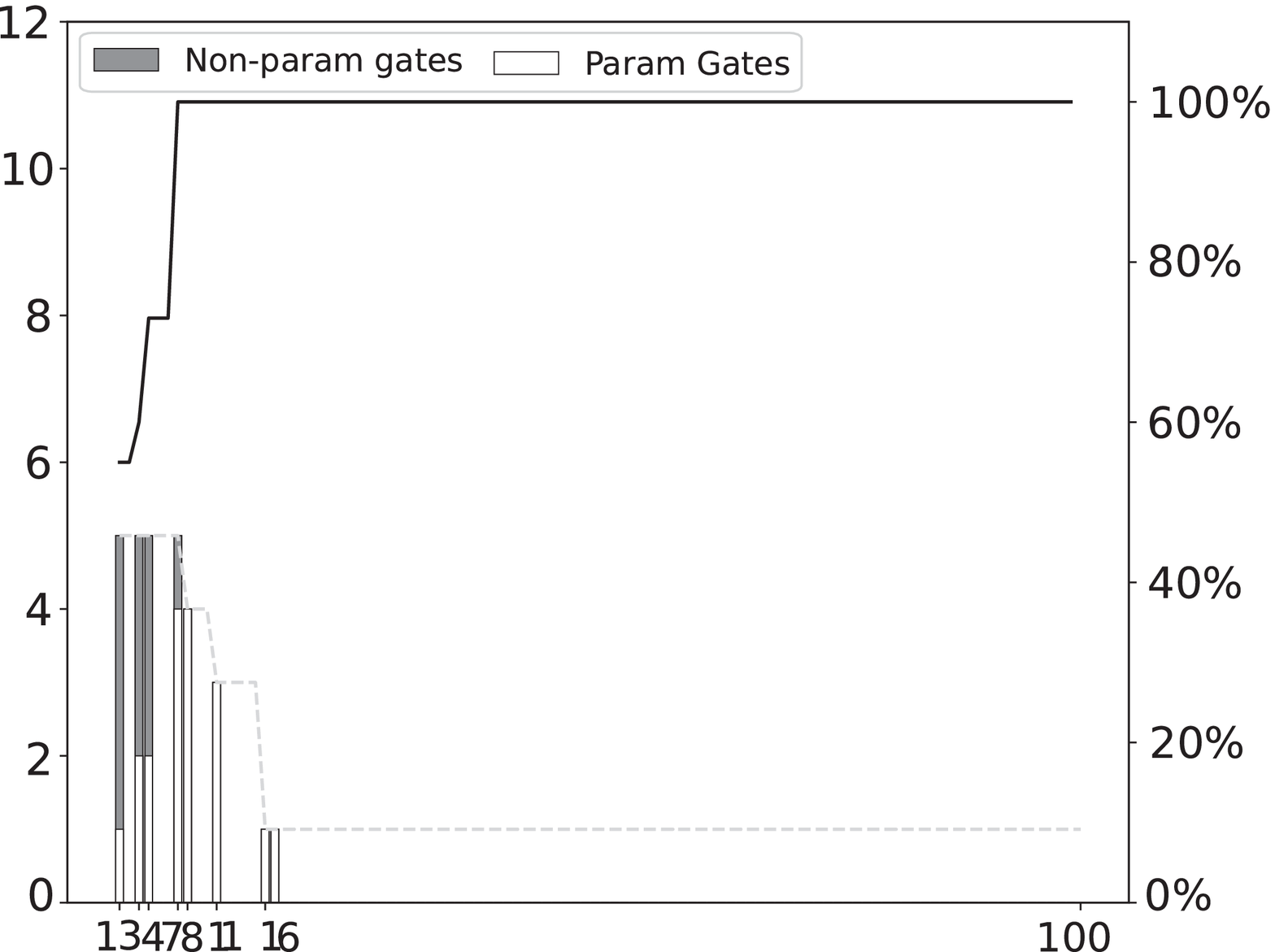}}
	\subfigure[] {\includegraphics[width=.31\textwidth]{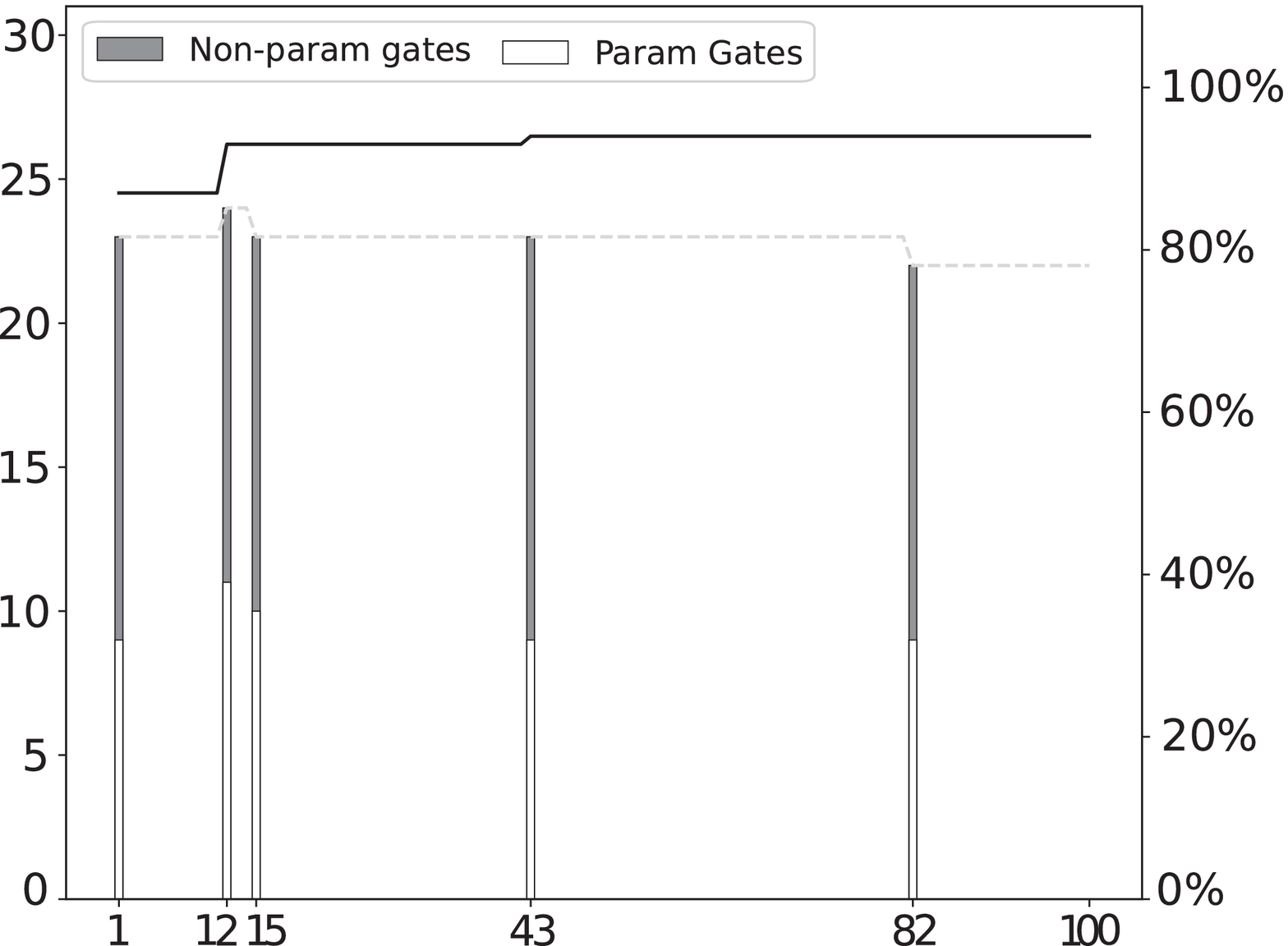}}
	\subfigure[] {\includegraphics[width=.305\textwidth]{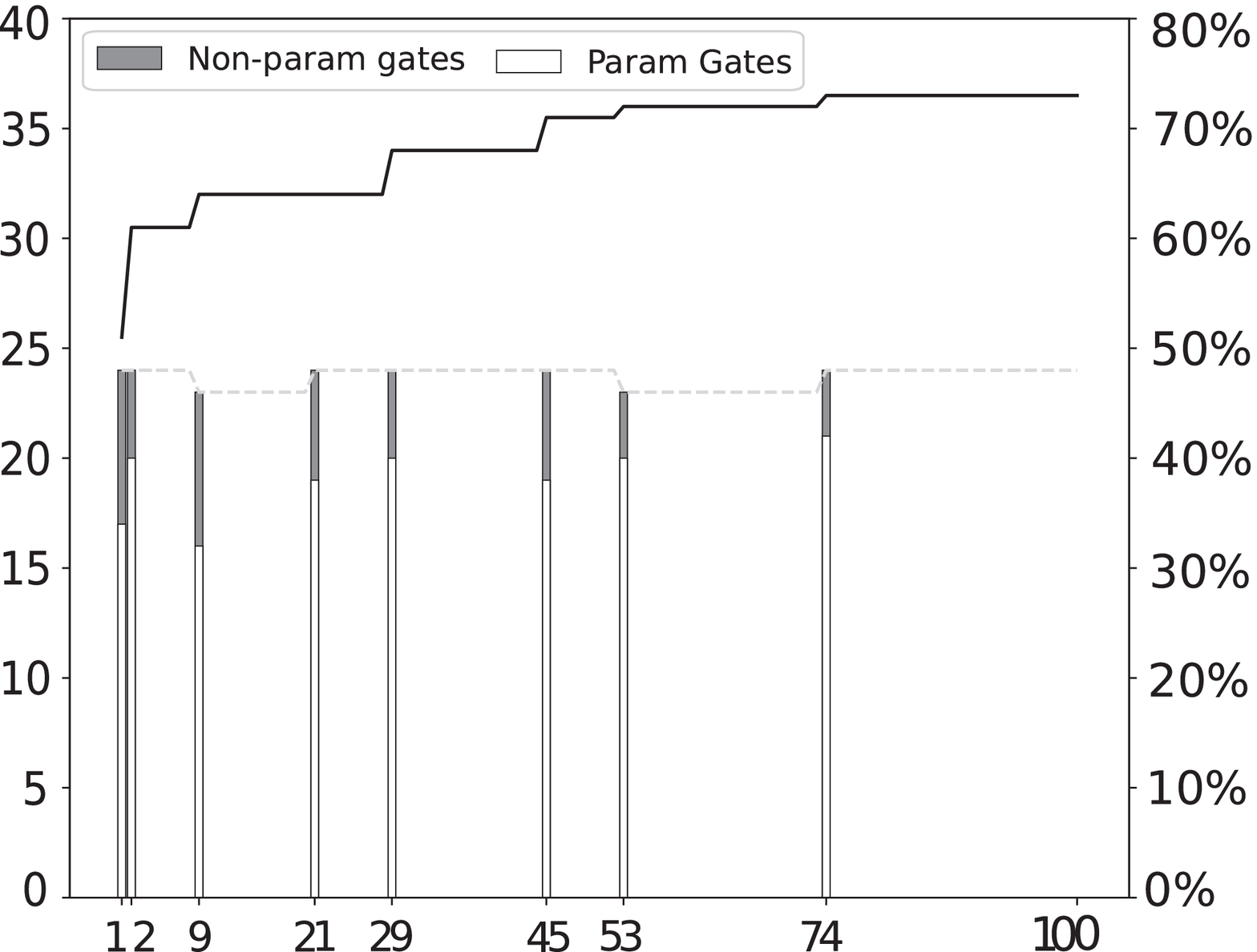}}
	\caption{The performances of the EQAS scheme for three datasets, Iris, MNIST and Fashion-MNIST.
		\label{fig:4}}
\end{figure*}

FIG.~\ref{fig:4} shows the accuracy(right axis) and the number of quantum gates(left axis) used of the searched QCA in every iteration on 3 different datasets, Iris, MNIST, and Fashon-MNIST, where FIG.~\ref{fig:4}(a) shows the searched processes of the classification task for the Iris dataset, FIG.~\ref{fig:4}(b) for the MNIST dataset, and FIG.~\ref{fig:4}(c) for the Fashion-MNIST dataset. The heights of bars are the number of gates used in the searched QCA, where the light color of bars shows the number of parameterized gates and the dark color shows the number of gates without parameters. The solid line shows the change in the accuracy of different QCAs during the searching processes in each iteration. The dashed line is the change of quantum gates used in the searched QCA and the QCA stays the same if the dashed line is horizontal and without bars. The numbers on the horizontal axis indicate that the QCA has changed at the iteration, that is, a QCA with better performance has been searched.
	
FIG.~\ref{fig:4}(a) shows that the proposed EQAS scheme obtains a searched QCA with 1 non-parameterized gate and 4 parameterized gates and achieves 100\% accuracy after 7 iterations. After 16 iterations, the 2-qubit QCA composed of only one parametrized quantum gate is found without reducing the accuracy.

FIG.~\ref{fig:4}(b) shows that the EQAS scheme needs more iterations to search for a QCA with fewer quantum gates and higher accuracy for the MNIST dataset. To achieve higher accuracy, the scheme allows the number of quantum gates to be increased. For example, the QCA reduced one non-parameterized quantum gate but increased two parameterized quantum gates compared with the previous QCA at the 12th iteration. Meanwhile, the EQAS scheme can search for QCA with fewer quantum gates while maintaining accuracy or even improving accuracies, such as at the 43rd iteration and the 82nd iteration. Finally, the searched 4-qubit QCA for the MNIST dataset achieves 94\% accuracy with 22 quantum gates which have 9 parameterized gates.

FIG.~\ref{fig:4}(c) shows that the EQAS scheme can search QCAs with higher accuracy through evolution many times. Moreover, the number of parameterized quantum gates increases with the process of searching, indicating that the searched QCA needs more parameterized quantum gates to achieve higher accuracy in the classification task with the Fashion-MNIST dataset. Furthermore, the accuracy of the 4-qubit QCA obtained by the EQAS scheme after 100 iterations is 75\%.

\bibliography{mybibliography}
\end{document}